# Proximity-induced quasi-one-dimensional superconducting quantum anomalous Hall state: a promising scalable top-down approach towards localized Majorana modes

Omargeldi Atanov[1], Wai Ting Tai[1], Ying-Ming Xie[1], Yat Hei Ng[1], Molly A. Hammond[1], Tin Seng Manfred Ho[1], Tsin Hei Koo[1], Hui Li[1], Sui Lun Ho[1], Jian Lyu[1], Sukong Chong[2], Peng Zhang[2], Lixuan Tai[2], Jiannong Wang[1], Kam Tuen Law[1], Kang L. Wang[2,3,4], and Rolf Lortz[1]

[1]Department of Physics, The Hong Kong University of Science & Technology, Clear Water Bay, Kowloon, Hong Kong
[2]Department of Electrical and Computer Engineering, University of California, Los Angeles, CA 90095
[3]Department of Physics, University of California, Los Angeles, CA 90095
[4]Department of Materials Science and Engineering, University of California, Los Angeles, CA 90095

Abstract: In this work, ~100 nm wide quantum anomalous Hall insulator (QAHI) nanoribbons are etched from a two-dimensional QAHI film. One part of the nanoribbon is covered with superconducting Nb, while the other part is connected to an Au lead via two-dimensional QAHI regions. Andreev reflection spectroscopy measurements were performed, and multiple in-gap conductance peaks were observed in three different devices. In the presence of an increasing magnetic field perpendicular to the QAHI film, the multiple in-gap peak structure evolves into a single zero-bias conductance peak (ZBCP). Theoretical simulations suggest that the measurements are consistent with the scenario that the increasing magnetic field drives the nanoribbons from a multi-channel occupied regime to a single channel occupied regime, and that the ZBCP may be induced by zero energy Majorana modes as previously predicted [24]. Although further experiments are needed to clarify the nature of the ZBCP, we provide initial evidence that quasi-1D QAHI nanoribbon/superconductor heterostructures are new and promising platforms for realizing zero-energy Majorana modes.

*Introduction* - Among all superconductors (SC), topological SC are of particular technological significance due to the predicted formation of zero-energy Majorana bound states (MBS) [1]. These anyonic quasiparticles, which obey non-Abelian statistics, are of great interest for the realization of fault-tolerant quantum computation [2,3]. The study of MBS in condensed matter systems was triggered by proposals that p-wave superconductors can host MBS [4,5]. Later, Fu & Kane [6] predicted the presence of MBS in the vortex cores of heterostructures between an s-wave SC and a topological insulator, followed by a large number of theoretical works on this topic [7]. Examples of platforms that host MBS include proximity-coupled SC/semiconductor heterostructures [8-12], SC/topological insulator heterostructures [13], HgTe-based topological Josephson junctions [14], Fe atomic chains on a SC [15], proximity-coupled helical hinge states of Bi(111) films decorated with Fe clusters [16], magnetic island covered superconducting gold surfaces [17,18] and two-dimensional QAHI/SC heterostructures [19,20]. Despite the observation of some predicted physical properties of MBS, the existence of MBS is generally debated [21,22].

One of the promising approaches to engineering topological superconductors with MBS is based on placing semiconducting nanowire structures in the proximity to an s-wave SC [9,10]. However, it has been shown to be particularly difficult to distinguish MBS from other fermionic in-gap states due to disorder effects that push finite energy Andreev bound states to near-zero energy [10,23]. Furthermore, the conducting channels in semiconductor nanowires can be easily localized by disorder, and it is difficult to achieve a single-channel conducting regime. Here, we present a different top-down approach starting from epitaxially grown 2D heterostructures consisting of a QAHI and the s-wave SC niobium etched into quasi-1D nanoribbons as proposed in Ref. 24 (Fig. 1a). An important advantage of this quasi-1D structure is that one of the conducting channels near the bulk gap can arise from the topologically protected chiral edge states (see Fig. 1b&c), which is more robust against the localization effects of disorder. In addition, the quasi-1D nanoribbons are fabricated from two-dimensional QAHI films, which are insulating in the bulk, so that the chemical potential of the nanoribbons is naturally close to the bulk gap of the QAHI. As a result, even without gating very few conducting channels are occupied. These two properties make quasi-1D QAHI/SC heterostructures promising platforms for the realization of MBS. Importantly, many quasi-1D nanoribbons can be etched from a single QAHI film, making this platform easily scalable for the creation of a large number of MBS for the construction of quantum qubits [24]. Recently, it has been demonstrated that the quantum anomalous Hall effect persists in a Hall bar device with a width of only ~72 nm, demonstrating that a single topological conducting channel can indeed be realized across such narrow nanoribbons [25], as expected in Ref. 24.

In this Letter, we present tunnelling spectroscopy measurements of quasi-1D QAHI/SC heterostructures, as schematically shown in Fig. 1a. At zero magnetic field, a ZBCP and multiple in-gap tunneling peaks were observed (Fig. 2). Interestingly, when a magnetic field perpendicular to the QAHI plane is applied, the multiple peak structure evolves into a single ZBCP (Fig. 3a&b). Theoretically, we show that the observed tunneling behavior is consistent with the evolution from a multi-channel occupied superconducting nanowire to a single channel occupied superconducting nanowire as the magnetic field increases, as demonstrated by simulations (Fig. 3c). Therefore, it is indeed possible that the observed ZBCP can be associated with MBS. This work experimentally demonstrates that quasi-1D QAHI/SC is a promising platform for the realization of MBS. However, due to the complications of the experimental setup, such as the presence of magnetic domains, further experiments that can investigate the electrical gating dependence of the tunneling spectrum at fixed magnetic field are needed to confirm the MBS origin of the ZBCP.

*Quasi-1D QAH/SC heterostructures-* A QAHI is a topological state of matter that exhibits dissipation-free chiral edge modes in the absence of external magnetic fields [26,27]. It has been predicted that the counter-propagating chiral edge modes hybridize as their spacing approaches about 100 nm, leading to the opening of a hybridization gap and the emergence of a single helical conduction channel along the nanoribbon when the chemical potential is in the bulk gap of the 2D QAHI [24]. It has been predicted that such a quasi-1D system will feature a particularly large energy window in which only one single propagating helical channel is partially filled. When only a single helical channel is partially occupied and the nanowire is in proximity to a SC, localized zero energy MBS are expected to form at the ends of the nanowire.

Here, we use a focused ion beam milling technique to fabricate 1D nanoribbons joining a bare QAHI to a QAHI/SC heterostructure as schematically shown in Fig. 1d. The corresponding field emission scanning electron microscope image of Device 1 is shown in Fig. 1e. Overall, a 2D QAHI film is divided into several sections by focused ion beam (FIB) cuts as illustrated in Fig. 1e. The upper section (grey) is covered by Nb, which is a superconductor. The bottom section is covered by an Au electrode (orange). On the other hand, the middle section (blue) is separated into three electrically isolated regions by FIB cuts. It is important to note that the FIB cuts produce nanoribbons with a width of about 100 nm. A magnified view of a 135 nm wide nanoribbon is shown in Fig. 1f. The key is that part of the nanoribbon is covered by Nb, while the other part is connected to the Au lead via a two-dimensional QAHI with a typical width in the order of $100 \mu$m. Therefore, each section defined by the FIB cuts resembles the structure illustrated in Fig. 1a. When an odd number of conducting channels are occupied in a nanoribbon, a Majorana mode is expected to form near the interface between the Nb-covered nanoribbon and the bare nanoribbon [24]. A Majorana mode is schematically indicated as a red dot in Fig. 1a. Since the bare nanoribbon is conducting, the Majorana mode is indeed delocalized. However, the presence of the Majorana mode will induce resonant Andreev reflection at zero bias [24,28].
Therefore, devices such as the one shown in Fig. 1d allow the search for MBS using Andreev reflection experiments. It is expected that electrons from the Au lead can be injected into the nanowire through the QAHI (the blue region) and get Andreev reflected at the superconducting section of the nanowire. The presence of an MBS will result in a zero bias conductance peak. The first results from our devices are presented in Fig. 2, providing evidence for proximity induced superconductivity in ~100 nm wide nanoribbons. Such devices exhibit complex multi-peak structures in the differential conductance versus bias voltage with a pronounced ZBCP as mentioned above (Fig. 3a&b). By comparison with theoretical simulations, we attribute the conductance peaks to originating from the multiple conducting channels of the quasi-1D QAHI. Importantly, we demonstrate that the band structure can be tuned by an applied magnetic field. When the magnetic field is sufficiently large, the helical channel appears to be the only conducting channel (Fig. 3c). The differential conductance then exhibits a single sharp ZBCP as shown in Fig. 3a&b, which is similar to the theoretical prediction of a quasi-1D topological superconducting state in a QAHI [24]. In this paper, we report the Andreev reflection spectroscopy data of three different devices. The details of the device fabrication are discussed below.

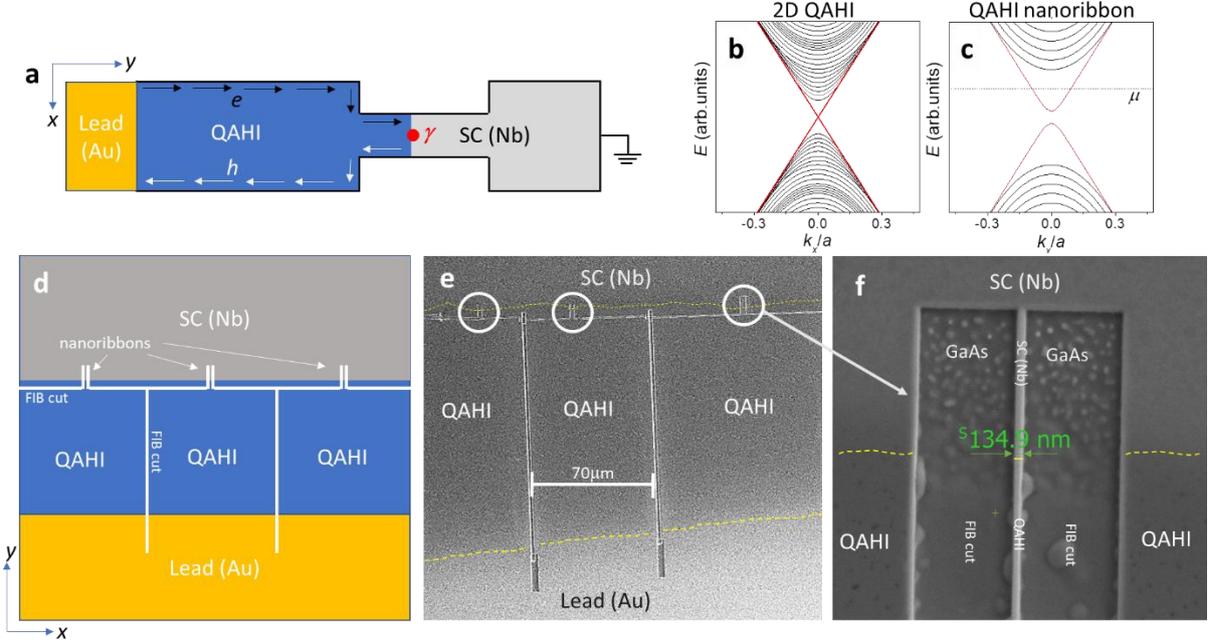

**FIG. 1.** (a) The two-terminal device proposed in ref. 24 for the detection of Majorana zero modes at a narrow 1D QAHI /QAHI-SC heterostructure junction, where an electron (black arrows) is reflected as a hole (white arrows) by a Majorana mode $\gamma$. (b) & (c) Qualitative illustration of the difference in band structure between a 2D QAHI with its chiral topological surface state (red) and a quasi-1D QAHI, respectively, where the hybridization of the chiral edge modes forms a helical channel and opens a small hybridization gap [24]. When the chemical potential $\mu$ cuts through the helical channel, there is only a single conducting channel in the nanowire. Sketch (d), and field emission scanning electron microscope image (e) of a QAHI nanoribbon device with multiple parallel nanoribbons fabricated by our focused ion beam (FIB) milling technique. The white circles in (e) mark the positions of the nanoribbons. The upper region is covered by the superconductor niobium and the nanoribbons extend from the bare QAHI region below into the niobium covered region. (f) Magnified view of a 135 nm wide nanoribbon. In the regions marked as GaAs, the substrate is expoded by the ion beam milling, leaving behind a narrow nanoribbon in between them, with the lower half of the QAHI nanowire connected to a 2D QAHI region, while the upper half is covered by Nb.

*Experimental Methods* - Single crystalline Cr-doped $(Cr_{0.12}Bi_{0.26}Sb_{0.62})_2Te_3$ QAHI films of 6 nm thickness were epitaxially grown on semi-insulating GaAs (111) substrates in an ultra-high vacuum molecular beam epitaxy system at UCLA [26,27,29,30]. The growth dynamics were monitored by surface sensitive reflection high energy electron diffraction (RHEED). The quality and composition of the films were characterized using standard techniques such as energy dispersive X-ray (EDX) spectrometry. Further details on the growth and characterization can be found elsewhere [27,29,30]. Subsequently, ~0.5 to 1 mm wide niobium strips of ~100 nm thickness were sputtered onto the QAHI layer using a shadow mask.
The QAHI – QAHI/SC nanostructures were fabricated at HKUST by cutting the 2D QAHI film using a FEI Helios G4 UX dual-beam focused ion beam (FIB) / field emission scanning electron microscope (FESEM) system. The typical cutting depth was 300 - 400 nm. The cutting process was monitored by the integrated FESEM. Our previous tests on larger 2D QAHI films have shown that the FIB technique does not affect the chiral QAHI edge mode, which appears to be very robust in these materials. The electrical connections were prepared by sputtering a thin Au layer on the QAHI surface prior to FIB treatment and connecting 25 μm thick Au wires with conductive Ag paint on the relatively large Au terminals to avoid contamination near the nanoribbons (Device 1), or with Al wires bonded by an ASM AB520 wedge bonder for all subsequent devices. The experiments were performed in a $^3$He cryostat with a base temperature of 300 mK equipped with a 15 T superconducting magnet. For Device 2, we preformed some measurements in the dilution refrigerator inset of a Quantum Design Physical Properties Measurement System. Fig. 1d shows a sketch of a typical device containing several nanoribbons in parallel and Fig. 1e is the corresponding field emission scanning electron microscope image of Device 2. The lower orange region is a lead, represented by an Au film evaporated onto the 2D QAHI surface. The blue area in the middle is an uncovered 2D QAHI area which is etched into separate areas by narrow grooves (white lines) cut all the way into the substrate by the FIB technique. The grey area at the top is the niobium-covered QAHI, separated from the blue uncovered QAHI region by another groove, except for 3 narrow nanoribbons that extend from the uncovered QAHI region across the boundary to the niobium covered region. Fig. 1f shows an enlarged view of one of the nanoribbons, formed between two regions where the GaAs substrate is exposed by the FIB milling.

Andreev reflection spectroscopy measurements were performed at HKUST using a Keithley 6221 AC/DC current source, a 34411A digital multimeter, and an SR830 lock-in amplifier. The current source generated an AC current with variable DC offset. The DC component induces the bias voltage by injecting the current in small steps from a negative value to a positive value across the point contact formed by the junction between the bare QAHI region to the QAHI/SC region within the nanoribbon. We used a current-stabilized technique because it allows a four-terminal configuration and provides better results for high transparency tunneling contacts than a voltage-stabilized mode. The AC current of constant amplitude comparable to the step size of the bias current scan was used to determine the differential conductance d$I$/d$V$ as a function of the measured bias voltage. The DC bias voltage was measured with the digital multimeter. The differential conductance d$I$/d$V$ can be derived either from the AC signal or by differentiating the DC signal. In some cases, the DC signal proved to be of superior quality (Device 1&2), while in general the AC technique provides a higher relative resolution. Devices 1 and 2 were measured in a two-terminal configuration as shown in Fig. 1d. The separately determined contact series resistance required a correction of the measured voltage. During the measurements, an unwanted parallel contact was detected in Device 1 and 2, caused by contamination of the sample edge with a very thin layer of Au. The ohmic contribution of this parallel contact was carefully determined separately and removed. It caused a constant positive offset in the conductance without affecting the size and shape of the Andreev signal. Device 1 consisted of 2 parallel 140 nm wide nanoribbons. Device 2 was fabricated with 10 parallel nanoribbons, but it was found that only one nanoribbon was intact and contributing to the conductivity. Device 3 was measured in a quasi-four-terminal configuration with an additional Au terminal on the bare QAHI side and two contacts on the niobium for current injection and voltage measurement respectively. It consisted of 9 parallel nanoribbons about 100 nm wide between the bare QAHI region on one side and the QAHI/SC heterostructure on the other. The many parallel nanoribbons, which are likely to contribute slightly different characteristics, resulted in relatively broad experimental features in applied magnetic fields. Due to its more precise four-terminal nature, this device served us as a test of the validity of the corrections made in the data analysis of Device 1 & 2 mentioned above (see Supplementary Information [31]).

*Experimental Results and Data Analysis* - Fig. 2. shows differential conductance data taken at ~300 mK as a function of bias voltage data obtained from the three different devices. The data in a & c were taken in zero field, while for Device 2 we applied a weak magnetic field of 0.3 T, as the zero-field data showed quite different behavior. We attribute the unusual zero-field data to domain formations that can be ordered by the weak magnetic field (see Supplementary Information [31]). All devices show an Andreev reflection peak approximately within the energy range of the Nb gap, with multiple shoulder and peak-like structures, most notably with a pronounced ZBCP. The overall Andreev reflection peaks of Devices 1 and 2 are of approximately the same magnitude with similar background conductance. For Device 3 with data taken at a lower temperature of 50 mK, the overall conductance is significantly smaller. Note that Device 3 consists of nine parallel nanoribbons, which may lead to some broadening effects due to the averaging of the contributions from different nanoribbons.

At first sight, the data are quite different from the predicted differential conductance at a quasi-1D

QAHI - QAHI/SC heterostructure junction [24], where the conductance was restricted to sharp, quantized ZBCPs. To characterize the multiple peak and shoulder structures and to obtain an estimate of the superconducting energy scale involved, we first analyze the data by fitting with a modified 1D Blonder-Tinkham-Klapwijk (BTK) model considering partially transparent tunnel junctions [32]. Fitting the fundamental features requires at least a three-gap s-wave pairing model. The fit can reproduce the features well with $\Delta_1$=0.16 meV, $\Delta_2$=1.14 meV, $\Delta_3$=1.4 meV and $\Delta_4$=2.2 meV, for Device 1 (Fig. 2a); $\Delta_1$=0.2 meV, $\Delta_2$=0.9 meV and $\Delta_3$=1.9 meV for Device 2 (Fig. 2b) and $\Delta_1$=0.16 meV, $\Delta_2$=1.25 meV and $\Delta_3$=2.5 meV for Device 3 (Fig. 2c). Note that the broader nature of the Andreev signal does not allow us to distinguish a 4$^{th}$ gap $\Delta_4$ for Sample 2 & 3, which, if present, is merged with $\Delta_3$ here. Considering this, very similar gap values are found in all devices, demonstrating the similarity of the physical origin of these structures, and proving the reproducibility of the results. The multi-gap structure suggests the contribution of multiple conducting channels, including the bulk channels of the QAHI nanoribbons. This is further evidenced by the increased background conductance of ~ 2 - 4 $e^2/h$ at high bias voltage in all devices. This is likely due to the nanostructuring which shifts the chemical potential of the heterostructure into the bulk bands. We will later show with detailed theoretical simulations that the multiple shoulders are indeed due to the contribution of bulk bands in addition to the topological channel, but this requires an analysis that goes far beyond the BTK model. It is noteworthy that the main features in the spectra fall in the range of the superconducting gap of the Nb film [31], which is known to be in between ~1.6 meV [20] and 2.3 meV [33].

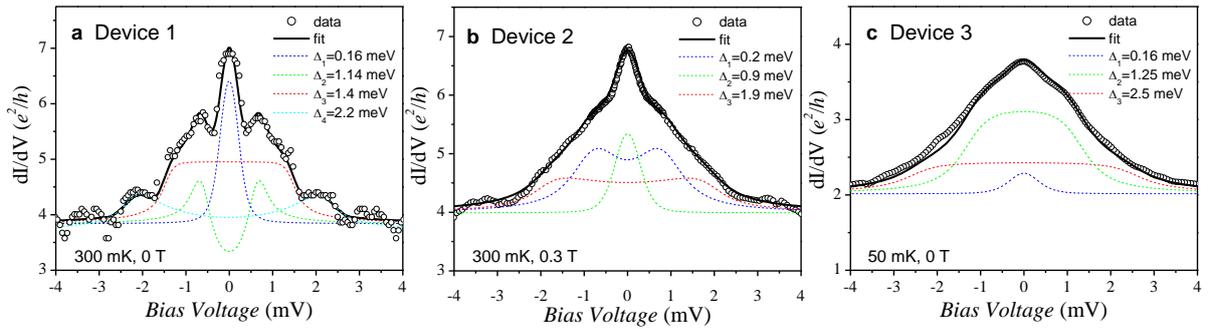

**FIG. 2.** a) Zero field Andreev reflection spectroscopy d$I$/d$V$ data (circles) measured in zero field at $T$ = 300 mK at the interface between a quasi-1D bare quantum anomalous Hall insulator (QAHI) and a QAHI / SC heterostructure within 140 nm wide nanoribbons (Device 1). The device consisted of 2 parallel nanoribbons and the data are normalized as d$I$/d$V$ per nanoribbon. The dashed lines are used to highlight the multiple peak and shoulder structures using a fit with a modified Blonder-Tinkham-Klapwijk (BTK) model, which assumes four superconducting gaps of $\Delta_1$=0.16 meV, $\Delta_2$=1.14 meV, $\Delta_3$=1.4 meV and $\Delta_3$=2.2 meV (see text for details). b) Similar data for Device 2 taken at 300 mK in a 0.3 T magnetic field. The BTK fit considers 3 gaps of $\Delta_1$=0.2 meV, $\Delta_2$=0.9 meV and $\Delta_3$=1.9 meV. c) Similar data for Device 3 taken in zero field at $T$ = 50 mK. A BTK fit considers 3 gaps of of $\Delta_1$=0.16 meV, $\Delta_2$=1.25 meV and $\Delta_3$=2.5 meV. The device consisted of 9 parallel nanowires. Data are normalized as d$I$/d$V$ per nanoribbon.

Fig. 3a shows data from Device 1 in different magnetic fields applied perpendicular to the basal plane of the QAHI and thus perpendicular to the orientation of the nanoribbon. It shows a rapid evolution of the contributions from the two gaps and the ZBCP, and a sudden reduction in the background conductance between 2.5 and 3.5 T. Up to 2 T, the overall Andreev peak narrows rapidly. At 3 T, $\Delta_2$ to $\Delta_4$ appear to be suppressed, leaving only the ZBCP ("$\Delta_1$"), which sharpens dramatically, and the background conductance begins to decrease, first at high bias and then with increasing field at lower bias voltages, until at 3.5 T the Andreev signal is reduced to a very sharp ZBCP. Finally, at 4 T, the data become completely flat.

The vanishing of the structures that we previously referred to as $\Delta_2$ to $\Delta_4$, followed by the rapid suppression of the background conductance between 2.5 T and 3.5 T, can be explained by the suppression of the relevant conducting channels in the nanoribbon. At zero field, the relatively high differential conductance indicates that the chemical potential $\mu$ is in the energy range of the bulk states of the QAHI (see dashed line labelled $\mu_5$ in the inset of Fig. 3d). As the magnetic field increases, it shifts relatively downward ($\mu_4$ & $\mu_3$) until it leaves the bulk states ($\mu_2$), causing the background conductance to drop to a much lower value. The disappearance of the bulk contribution causes the vanishing of $\Delta_2$ to $\Delta_4$. What remains is a sharp ZBCP at 3.25 T and 3.5 T, which we attribute to superconductivity in a single conducting channel ($\mu_2$), before the chemical potential shifts into the small hybridization gap opened up by the quantum confinement in the narrow nanoribbon ($\mu_1$). Note that the pairing persists in such narrow Nb strips covering the QAHI nanoribbons up to almost 8 T, as evidenced in a separate experiment (Suppl. Fig. 1a in Ref. 31). Therefore, a change in the number of channels as a function of the applied magnetic field could provide a plausible explanation for the observed evolution of the Andreev spectra. At present, the exact mechanism by which the magnetic field can tune the number of channels in this complicated setup is still unclear. However, it is plausible that an external magnetic field can increase the net magnetization of the nanoribbon and increase the bulk gap, so that the bulk channels are pushed above the Fermi energy as the field increases. In the Supplementary Information [31] we demonstrate this with simulated band structure and d$I$/d$V$ data for 3 different magnetization values (Suppl. Fig. 8 & 9). In either case (fixed band structure with variable $\mu$ as assumed in Fig. 3c&d, or variable band structure with fixed $\mu$), qualitatively similar tunneling spectra can be simulated. Another reason for the (relative) shift of the chemical potential with respect to the bandstructure could be a magnetic field dependence of the coupling between the Nb film and the QAHI layer.

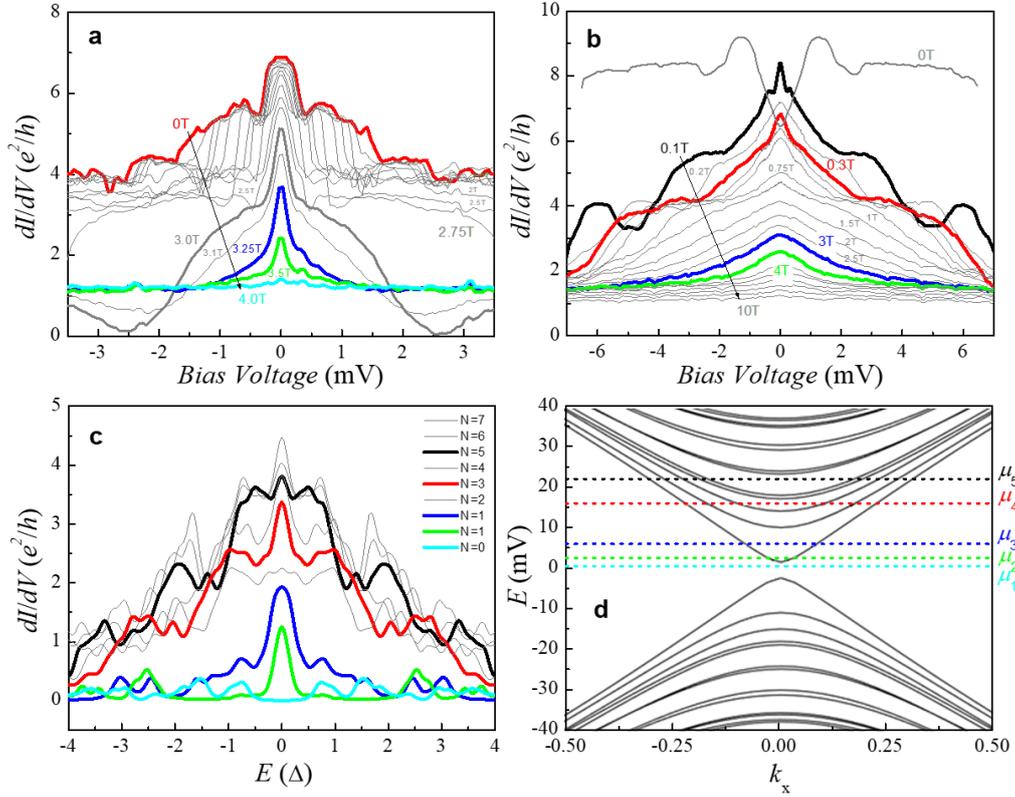

**FIG. 3.** a) Andreev reflection spectroscopy d$I$/d$V$ data (normalized per nanoribbon) as a function of bias voltage measured at the interface between a bare QAHI and a QAHI / SC heterostructure in Device 1 within 140 nm wide nanoribbons in various magnetic fields (0, 0.02, 0.04, 0.1, 0.2, 0.5, 1.0, 1.5, 2.0, 2.5, 2.75, 3.0, 3.1, 3.25, 3.5, 4.0T) applied perpendicular to the basal plane of the QAHI. The data show a complex evolution due to the shift of the chemical potential relative to the conduction band structure. This gradually suppresses the multiple peak and shoulder structures, until only a sharp zero-bias peak remains at 3.25 T and 3.5 T. b) Similar Andreev reflection spectroscopy d$I$/d$V$ data for Device 2 in various perpendicular magnetic fields (0, 0.1, 0.2, 0.3, 0.4, 0.5, 0.75, 1.0, 1.5, 2.0, 2.5, 3.0, 4, 5, …, 10T). c) The calculated conductance spectra (d$I$/d$V$ as a function of bias energy $E$, $\Delta$ is the pairing potential) for the geometry similar to Fig. 1a. $\mu$ denotes the chemical potential of the quasi-1D QAHI region, while N denotes the number of channels occupied by the corresponding chemical potential. (d) The corresponding energy spectrum of the quasi-1D QAHI of width 120 nm assumed in the calculation, where the dashed lines indicate the positions of the chemical potentials $\mu_5 = 22$ meV (black), $\mu_4 = 16$ meV (red), $\mu_3 = 6$ meV (blue), $\mu_2 = 2.5$ meV (green) and $\mu_1 = 0.5$ meV (cyan). The corresponding data in panels a-c are highlighted with the same colors.

Device 2, whose data are shown in Fig. 3b, shows a very similar behavior (except for the zero-field data, which features only a large constant background conductance with a V-shaped pairing gap). The high zero-field conductance indicates that the zero-field chemical potential of this device is well within the conduction band. Magnetic domains may also contribute to the unusual zero-field data. However, at 0.1 T, the shape is similar to the zero-field conductance of Device 1, although its magnitude is higher and the zero-bias peak sharper (indicating that the chemical potential is slightly higher and crosses more conduction bands). Several periodic conductance drops are observed at higher bias voltages beyond the pairing gap. At higher magnetic fields the conductance drops further, and at 0.3 T the data match the zero-field data of Device 1 in height and shape (see also Fig. 2). This suggests that the chemical potentials of Device 1 at zero field and Device 2 at 0.3 T are at a comparable energy level (e.g. with the same number of conducting channels occupied). At higher fields, the Andreev peak continues to fall in Device 2 in a similar manner to Device 1, except that the zero bias anomaly becomes much broader and fades into the background until the conductance falls to a constant value close to $e^2/h$ at 10 T. Further experimental data from Device 3 can be found in the Supplementary Information [31].

To further confirm the above analysis, we perform a numerical simulation in which we calculate the conductance spectra using a geometry similar to the experiment (see details in Ref. 31). The resulting conductance spectra (d$I$/d$V$ versus energy $E$) are shown in Fig. 3c, while Fig. 3d shows the corresponding band structure with five lines representing the cases with different chemical potentials ranging from $\mu_5$ = 22meV to $\mu_1$=0 meV (the colors mark the corresponding curves in Fig 3a–c). The simulated conductance curves reproduce the main features of the experimental data in Fig. 3a&b. For example, there are similar multiple peaks and shoulders in the calculated spectra. Due to the strong Andreev reflection, the spectra have a larger shoulder within the pairing gap. Importantly, there is a central peak around the zero-bias caused by low energy Andreev bound states in the gap. Beyond the pairing gap, the change in density of states in the bulk subbands causes a periodic drop in the conductance at large bias (Fig. 3c). As the chemical potential is gradually tuned to cut through only one channel (blue and green data associated with $\mu_3$ and $\mu_4$), the bulk contributions decrease, leaving a ZBCP caused by the zero energy Majorana mode, consistent with the measurements shown Fig. 3a &b.

*Conclusion* – In conclusion, the experimentally observed complex differential conductance spectra with multiple shoulder structures and a ZBCP, as well as the enhanced normal state conductance, suggest that nanostructuring can bring the chemical potential in the zero field close to the bulk bands, so that only a few conducting channels are occupied in the nanoribbon even without gating. By applying a magnetic field, the QAHI nanowires can possibly be tuned to the single channel limit. The data from Device 1 implies that this limit is reached at a field of 3.25 T, where the ZBCP takes on a much sharper and narrower shape. Theoretically, an MBS is then expected to cause resonant Andreev reflection with a sharp, quantized ZBCP [24]. Experimental confirmation of the latter will require extensive further careful testing [34] and detailed experiments at lower temperatures in improved devices, ideally with gate electrodes to precisely tune the chemical potential. However, the rich variety of unusual superconducting properties in this ferromagnetic topological insulator in the 1D limit underlines the high potential of this platform, which is fabricated with a relatively simple top-down approach starting from thin-film wafer technology.

*Rev. Lett.* **113**, 137201 (2014).

# Supplementary Information

# Proximity-induced quasi-one-dimensional superconducting quantum anomalous Hall state: a scalable top-down approach towards localized Majorana modes

by Atanov et al.

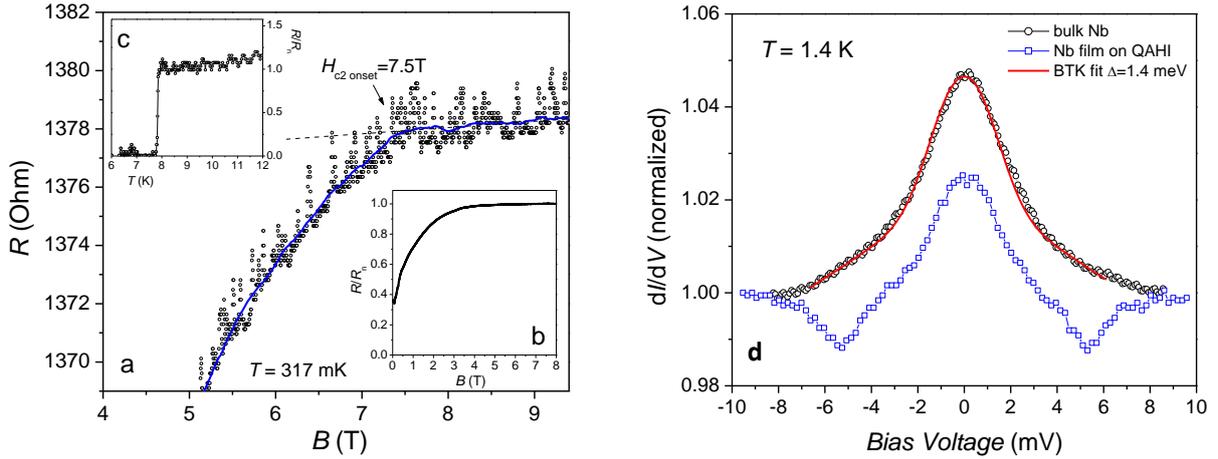

**Supplementary Figure 1.** a) Electrical magnetoresistance near the upper critical field measured at $T = 317$ mK across a ~100 nm side Nb/QAHI nanoribbon using a 3-terminal method. The contact resistance is included in the data and causes a positive offset. The arrow indicates the onset of the upper critical field $H_{c2} = 7.5$ T. Black circles represent the raw data while the blue line is smoothed over 250 points. b) Same magnetoresistance data (normalized by the normal state resistance $R_n$) shown at full scale. The finite resistance is due to the contact resistance and the contribution from a normal conducting region. c) Electrical resistance of a 2D Nb/QAHI film (normalized by $R_n$)) showing the superconducting transition at $T_c = 7.8$ K. d) Normalized Andreev reflection spectroscopy $dI/dV$ data measured with a scanning probe on a bulk Nb sample and on the surface of the Nb layer grown on the QAHI film. A BTK fit was added to the bulk data with a gap value $\Delta = 1.4$ meV. Note that these data were measured in a $^4$He cryostat at $T = 1.4$ K, which is higher than in our other experiments in a $^3$He cryostat. This reduces the gap value slightly. The additional shoulder and dip structures in the Nb/QAHI heterostructure are likely due to the QAHI band structure.

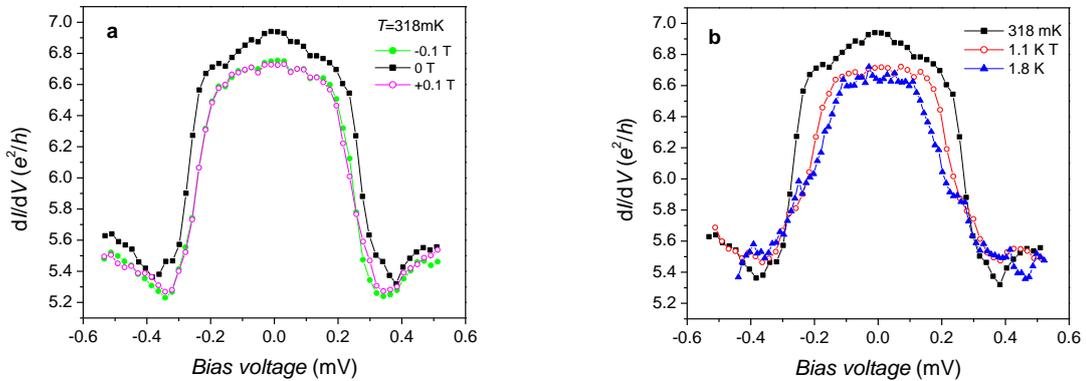

**Supplementary Figure 2.** Supplementary Andreev reflection spectroscopy $dI/dV$ data measured at the interface between a quasi-1D QAHI and a QAHI/SC heterostructure in Device 1. a) Details of the zero-bias peak in data taken at 318 mK in + and -0.1 T to illustrate the symmetry with respect to field reversal (black

squares are zero field data added for comparison). b) Details of the zero-bias peak in zero field at different temperatures.

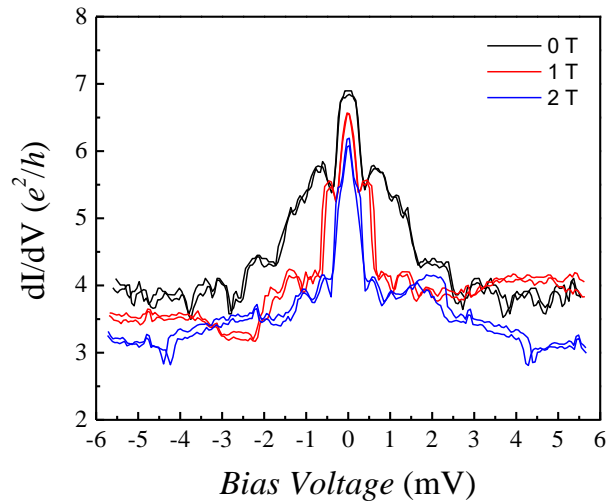

**Supplementary Figure 3.** Andreev reflection spectroscopy d$I$/d$V$ data measured at the interface between a quasi-1D QAHI and a QAHI/SC heterostructure in Device 1 at 0 T, 1 T and 2 T, with two data sets for each field taken on two different days, showing the reproducibility of even small details in repeated measurements.

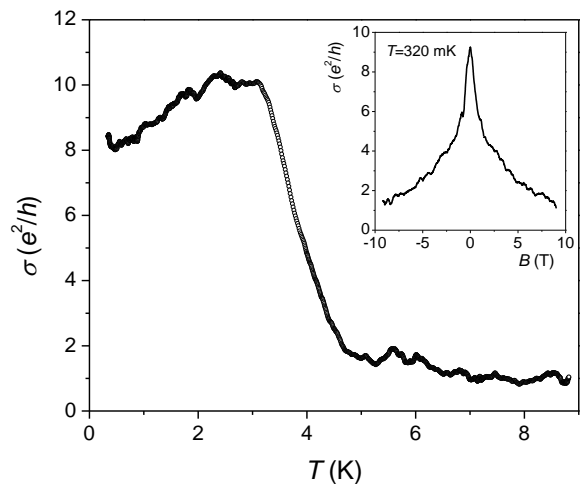

**Supplementary Figure 4.** Zero bias conductance measured as a function of temperature across the nanoribbon in Device 2 in zero magnetic field. It shows that the proximity induced superconductivity in the QAHI/SC heterostructure develops below 5 K, well below the superconducting transition of the niobium layer ($T_c$= 7.8 K). Inset: Zero-bias magneto-conductance measured across the nanoribbon in Device 2 as a function of the magnetic field applied perpendicular to the basal plane of the QAHI.

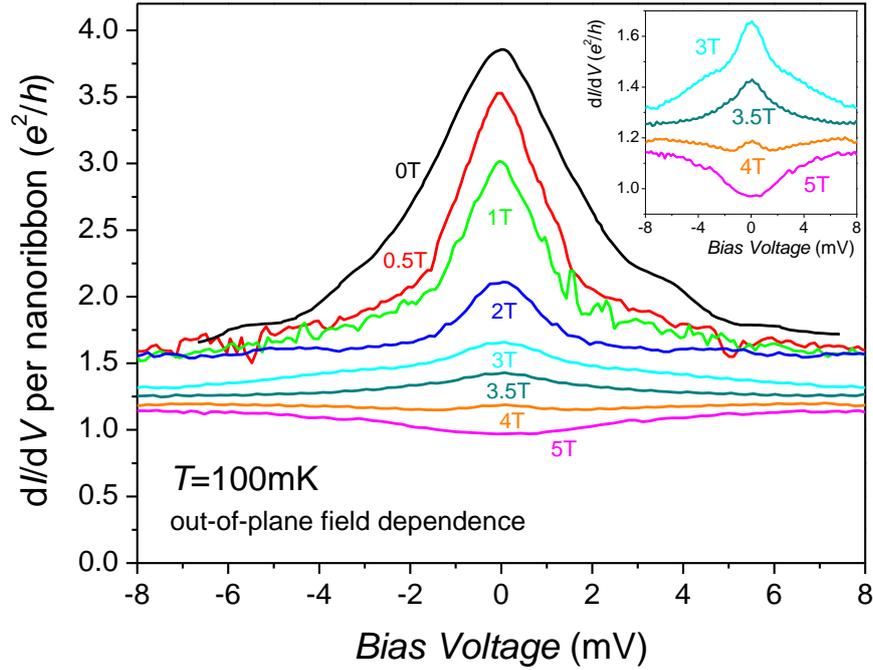

**Supplementary Figure 5.** Andreev reflection spectroscopy d$I$/d$V$ data measured at the interface between a quasi-1D QAHI and a QAHI / SC heterostructure in a device with 9 parallel nanoribbons (Device 3). The data were recorded at $T$ = 100 mK. The inset shows an enlarged view of the high-field data, showing the evolution of the zero-bias anomaly at higher fields in more detail.

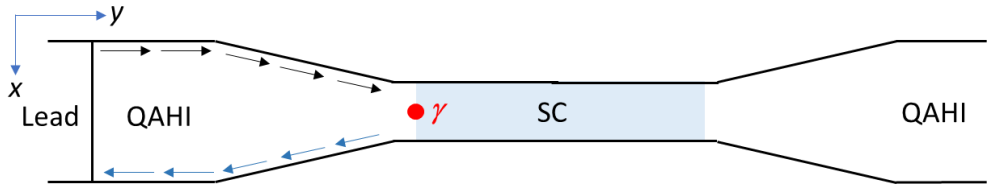

**Supplementary Figure 6.** Schematic representation of the device geometry adopted in the theoretical simulations for the d$I$/d$V$ spectra. In consistence with the experiment, we have created a nanoribbon region in the center of a wider QAHI sample. The superconductor is coupled to a part of the ribbon region. It illustrates how in our QAHI /QAHI-SC heterostructure 1D junctions an electron (black arrows) is reflected as a hole (blue arrows) by a Majorana mode $\gamma$.

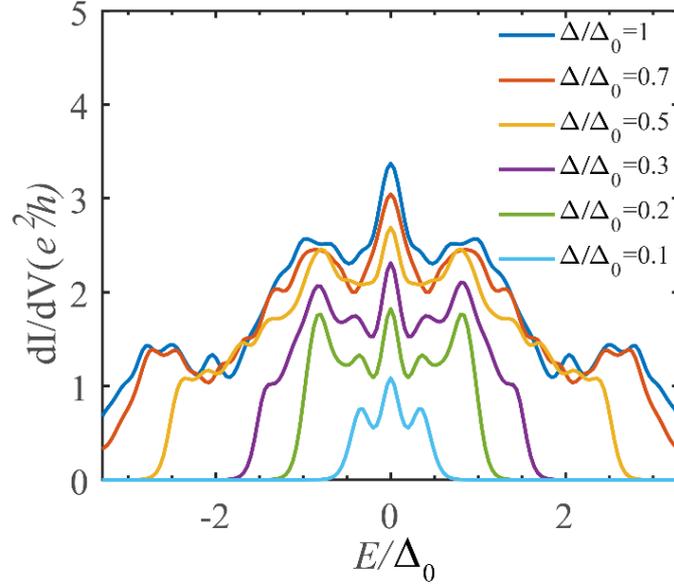

**Supplementary Figure 7.** Theoretical simulation of d$I$/d$V$ (in unit of $e^2/h$) as a function of bias energy $E$ for different pairing gaps $\Delta$ (in unit of $\Delta_0$= 2 meV). The shoulders of the conductance peaks become narrower when the pairing gap is reduced, similar to what is observed in Device 1 as a consequence of applied perpendicular magnetic fields between zero field and 2 T (Fig. 3 of the main article).

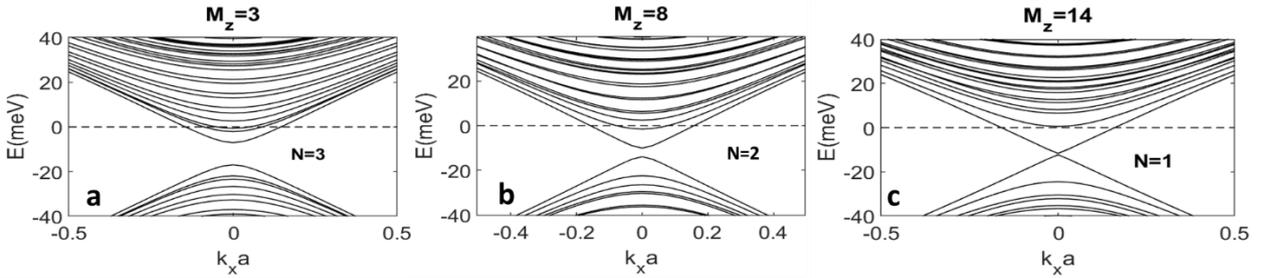

**Supplementary Figure 8.** Theoretical simulation of the band structure of a QAHI nanoribbon for three different values of the magnetization perpendicular to the basal plane of the QAHI. Here the magnetization energy (in units of meV) $M_z$ = 3 (a), 8 (b) and 14 (c), and the number of channels $N$ cut by the Fermi energy are highlighted

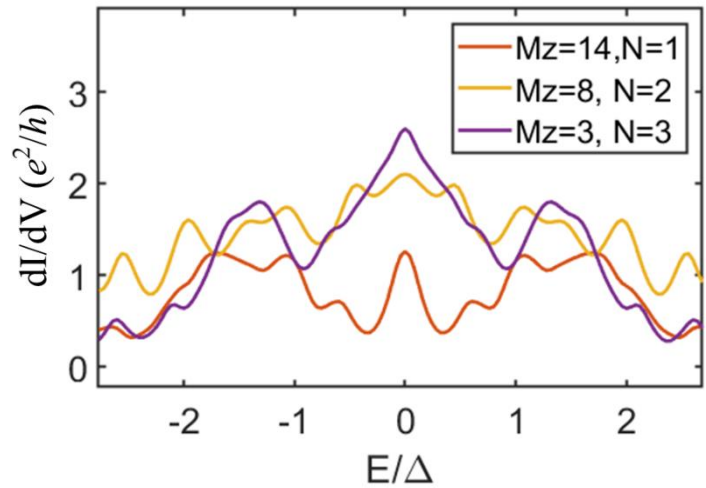

**Supplementary Figure 9.** Theoretical simulation of d$I$/d$V$ (in unit of $e^2/h$) as a function of bias energy $E$ at different values of the magnetization perpendicular to the basal plane of the QAHI, corresponding to the band structures shown in Suppl. Fig. 8.

## Supplementary Notes

### Overview of the different devices used in this study

**Device 1.** The device contained two ~140 nm wide nanoribbons that were measured in parallel. All data in this article are given as d$I$/d$V$ per nanoribbon. For technical reasons, a 2-terminal configuration was used and the contact resistance, which required a correction of the measured voltage, was determined separately. It was found that an unwanted parallel resistance of ~260 Ohms, resulting from a very thin gold contamination at the edge of the sample in contact with the leads on both sides, added a constant offset to the conductance, which was removed. We estimate that the accuracy of the corrected background conductance is limited to ~0.3$e^2$/$h$ due to this correction, while the magnitude and shape of the d$I$/d$V$ anomaly is unaffected as the conductances of the parallel resistors simply add up. The resolution was sufficient to observe small details of the conductance across the nanoribbons. The data were measured in DC mode as the resolution of the AC mode proved to be too noisy.

**Device 2.** This device contained 10 parallel nanoribbons of ~100 nm width. For technical reasons, a 2-terminal configuration was used and the contact resistance, which required a correction of the measured voltage, was determined separately. As the device was cut from the same wafer as Device 1, an unwanted parallel contact from a very thin gold layer contamination on one edge of the sample also provided a parallel contact of ~70 Ohms. This added a constant offset to the conductance, which was removed. We estimate that the accuracy of the corrected background conductance is limited by this correction to an order of 0.3$e^2$/$h$, while the magnitude and shape of the d$I$/d$V$ anomaly is unaffected, as the conductances of the parallel resistors simply add up. Comparison of the signal magnitude with Device 1 and 3 suggests that only one of the nanoribbons was continuous and contributing to the conductance.

**Device 3**. This device contained 9 parallel nanoribbons of ~100 nm width. All data is provided in d$I$/d$V$ per nanoribbon. All electrical transport measurements were made using a quasi-4-terminal method with two Au terminals deposited on the 2D QAHI area connected to the bare side of the QAHI nanoribbon and two contacts on the niobium layer for separate current injection and voltage measurement. All nanoribbons appeared to be of good quality and contributed in parallel to the total conductance. The 4-terminal method allowed us to use this sample to correct the effect of contact resistance on the bias voltage in the other devices. All data for this device were measured in both AC and DC mode, giving similar results. The data presented here were measured in DC mode.

### Discussion of the additional data of device 1 - 3

Suppl. Fig. 1a shows the electrical magnetoresistance of a ~100 nm wide Nb/QAHI nanoribbon near the upper critical field measured at $T$ = 317 mK using a 3-terminal method. The data include a positive offset from the contact resistance, which has not been corrected. The magnetoresistance shows a typical 1D behavior where current-induced phase slips cause a finite resistance with very continuous upper critical field transition. The data approach the almost constant normal state behavior at 7.5 T, confirming that pairing is present up to this field. Suppl. Fig. 1b shows the same data at full scale. Suppl. Fig. 1c shows an electrical resistance measurement made on a 2D part of our Nb-covered QAHI sample. The superconducting transition can be seen as a sharp jump initiating at 7.8 K. Note that the width of our 1D nanoribbon is comparable to the London penetration depth of Nb, which causes this unusually high upper critical field [1].

Suppl. Fig. 1b shows Andreev reflection d$I$/d$V$ data measured with a scanning probe at the surface of a bulk Nb sample compared to a QAHI/Nb heterostructure. The spectra were measured at $T$ = 1.4 K, which is significantly higher than in our $^3$He cryostat experiments. We have adjusted the contact transparency to obtain a similar Andreev reflection signal as in our nanoribbon experiments. Both spectra look qualitatively similar, with a similar width of the Andreev peak, which has a relatively broad shape, partly due to a limited quasiparticle lifetime as well as the elevated temperature. A fit of the BTK model to the bulk data provides a value for the superconducting gap of $\Delta$ = 1.4 meV. Note that this value is certainly somewhat lower than the gap value at the $^3$He temperatures at which our experiments were performed.

Suppl. Fig. 2 shows high-resolution d$I$/d$V$ data from Device 1 in a bias voltage range limited to the central zero-bias peak. The data reveal some substructure within the zero-field peak, which looks somewhat like a narrow plateau with a roof or an additional small peak in the center. Three sets of data are included in (a). First, the QAHI was fully magnetized in a field of -0.5 T, then we repeated the d$I$/d$V$ measurements in small magnetic field increments from negative to positive fields. No hysteresis was observed and the data in negative fields look identical to those in positive fields of the same magnitude, as shown here using -0.1 T and + 0.1 T data. Suppl. Fig. 2 b) shows the same zero field data at ~300 mK compared to data at 1.1 K and 1.8 K, showing a similar transition to a flat plateau-like top of the peak structure at elevated temperature and a reduction of the width of the gap as the proximity induced gap is gradually suppressed.

Suppl. Fig 3 presents repeated measurements of the d$I$/d$V$ data of Device 1 in zero field, 1T and 2 T magnetic fields applied perpendicular to the basal plane of the QAHI. The data were measured in two separate experiments on two consecutive days. The data illustrate the reproducibility of the data, with even some tiny spikes, dips or oscillatory structures appearing similar in both sets of data taken at the same field.

Suppl. Fig. 4 shows a measurement of the zero-bias conductance of Device 2 in zero field as a function of temperature. The Andreev signal from proximity-induced superconductivity in the QAHI is completely suppressed at temperatures above $T$ = 5 K, while the niobium film remains superconducting up to $T$ = 7.8 K (Fig. 1c). This shows that the observed Andreev reflection signals are indeed from a superconducting QAHI. The inset shows the zero-bias magneto-conductance, which is completely symmetric with respect to the field reversal, with a sharp peak at zero magnetic field.

Suppl. Fig. 5 shows d$I$/d$V$ data of Device 3 in various magnetic fields applied perpendicular to the basal plane of the QAHI measured at a temperature of 100 mK. In the energy range of the pairing gap, an Andreev reflection signal is observed which is qualitatively similar to that of Device 1, but the shoulder and peak structures are slightly broader. The broadening is simply a consequence of the parallel measurement of 9 nanoribbons, each with slightly different structures and slightly different bias voltages generated at the interface with the superconducting region, as a consequence of our current-driven measurement technique. However, the overall trend is very similar to Device 1, with an increased background conductance at low fields due to the presence of multiple channels when the chemical potential crosses the bulk conduction bands at low fields, which gradually decreases at higher fields. In the magnified view in the inset of Suppl. Fig. 5, a ZBCP is clearly distinguishable at fields above 2 T and disappears only at 4 T, which is very similar to the data from Device 1 at fields just above 3 T, although the ZBCP in Device 3 is wider due to the parallel nanoribbons. It is worth noting that Device 3 was measured using a quasi-4-terminal technique with no series contact resistance on the bare QAHI side, and the overall similarity confirms the corrections made in the other devices.

**Theoretical simulation models and methods**

*QAHI nanowire:* The QAHI nanoribbon with a finite width ($1 \leq n_y \leq N_y$) is described by:

$$H_{QAHI} = \sum_{k_x, 1 \leq n_y \leq N_y} \varphi^\dagger_{k_x,n_y}[(m_0 - 4m_1 + 2m_1\cos k_x)\tau_x - \hbar v_F \sin k_x \tau_z \sigma_y + M_z \sigma_z - \mu]\varphi_{k_x,n_y} + \varphi^\dagger_{k_x,n_y}(m_1\tau_x - \frac{i\hbar v_F}{2}\tau_z\sigma_x)\varphi_{k_x,n_y} + h.c..  \quad (1)$$

Here, the four-component electron annihilation operator $\varphi = [\varphi_{t\uparrow}, \varphi_{t\downarrow}, \varphi_{b\uparrow}, \varphi_{b\downarrow}]$ with t/b labels top/bottom and the ↑/↓ labels spin up/down, $k_x$ is the momentum along x-direction, $n_y$ is a site index, $\tau$ and $\sigma$ operator on layer and spin basis, respectively, $M_z$ is the magnetization energy, $\mu$ is the chemical potential, other terms capture the hybridizing bottom and top surface states in the quasi-one D limit. According to the experimental data of topological insulator surface states, we set the parameters as $\hbar v_F = 3$ eV·Å, $m_0 = -5$ meV, $m_1 = 15$ eV·Å. In practice, we have chosen a lattice constant $a = 4$ nm, which is sufficient to capture the low energy physics. Further details of the model can be found in references [3] & [4]. The band structures of the QAHI nanoribon shown in the main text and Suppl. Fig. 8 are calculated by diagonalizing the normal Hamiltnoian $H_{QAHI}$ with a width of 120 nm. Note that although we have illustrated the QAHI nanoribon model here by setting the nanoribon along x-direction, the energy spectrum would be the same if we set the nanoribbon along the other direction, due to the rotational symmetry in the origional 2D QAHI low-energy model [4].

*QAHI nanowire/SC heterostructure:* In the proximity of a niobium superconductor, it has been demonstrated that superconductivity can be induced in a QAHI [3]. The effective tight-binding model for the QAHI nanowire in this case can be described as follows:

$$H_{QAHI/SC} = \sum_{k_x, 1 \leq n_y \leq N_y} \psi^\dagger_{k_x,n_y}[(m_0 - 4m_1 + 2m_1\cos k_x)\tau_x s_z - \hbar v_F \sin k_x \tau_z \sigma_y s_z + M_z \sigma_z s_z + \Delta\frac{\tau_z+1}{2}\sigma_y s_y - \mu s_z]\psi_{k_x,n_y} + \psi^\dagger_{k_x,n_y}(m_1\tau_x s_z - \frac{i\hbar v_F}{2}\tau_z\sigma_x)\psi_{k_x,n_y} + h.c.. \quad (2)$$

Here, the Nambu basis $\psi = (\varphi, \varphi^\dagger)$, and the Pauli matrices $s$ operate in the particle-hole space. In line with with previous works [3], the superconducting pairing order parameter is added only for the top layer, which is set to be near the parent superconductor niobium.

*Calculation of the dI/dV spectra:* To be consistent with the experiment, we calculate the d$I$/d$V$ spectra a device geometry illustrated in Suppl. Fig. 6. With the above geometry and the constructed tight-binding models for the QAHI nanowire (note that it is straightforward to write the *y*-direction in real space by performing a Fourier transformation: $\psi_{n_x,n_y} = \frac{1}{\sqrt{N_x}}\sum_{k_x}\psi_{k_x,n_y}e^{-ik_x a}$), we can now calculate the differential conductance by using the scattering matrix approach

$$G_c(E) = \frac{e^2}{h}\text{Tr}[I - R_{ee}(E)R_{ee}^\dagger(E) + R_{he}R_{he}^\dagger(E)], \quad (3)$$

where $R_{ee}(E)$ and $R_{he}(E)$ represent the scattering matrices of the normal and the Andreev reflection contributions, respectively. The scattering matrices can be calculated numerically using the standard lattice Green's function method [2, 3].

To be specific, we set the ribbon region, as shown in Suppl. Fig. 6 to a width of 120 nm and the width at both ends to 200 nm, the length of the QAHI ribbon to 12 μm, the bare QAHI on the right side to 8 μm. The pairing gap in the experiment is 1 - 2 meV. To reduce the finite size effect, we adopted the zero-field pairing gap Δ to be 2 meV. It is also worth noting that the interface between the lead and the QAHI is very transparent. To simulate this transparency, we use a QAHI lead but fix the chemical potential deep in the bulk ($\mu = 40$ meV).

Further simulation data of d*I*/d*V* in units of $e^2/h$) versus *E* are shown in Suppl. Fig. 7 with different pairing gaps $\Delta$ (in units of $\Delta_0 = 2$ meV). The shoulders of the conductance peaks become narrower as the pairing gap is reduced, similar to what is observed in Device 1 as a consequence of the applied perpendicular magnetic fields between zero field and 2 T (see Fig. 3a of the main article).

Suppl. Fig. 8 shows theoretical simulations of the band structure of a QAHI nanoribbon with the device geometry described in Suppl. Fig. 6 for three different values of magnetization perpendicular to the basal plane of the QAHI: $M_z$ = 3 meV (a), 8 meV (b) and 14 meV (c). It can be seen that an increasing magnetization opens the bulk gap, which could be a possible plausible explanation for the magnetic field induced relative shift of the chemical potential observed in our experiments. The opening of the gap causes a relative shift of the chemical potential that crosses 3 bands in (a), 2 bands in (b) and only one band in (c).

Suppl. Fig. 9 shows the corresponding simulated differential conductance data for these three different $M_z$ values, showing the evolution from a state where the chemical potential crosses three bands (N=3) to a single band (N=1). While the high magnetization value of $M_z$ = 14 meV chosen here to simulate the N=3 state results in a somewhat different spectrum to the experimental zero field data from device 1, the simulated data are quite simlilar to the data of device 3 in Suppl. Fig. 5, with the N=3 state resembling the zero-field data and the N=1 state resembling the 4-T data in the inset of Suppl. Fig. 5, where a single peak also appears to be located in a dip within the bias region of the superconducting gap.

## Supplementary References